\documentclass[twocolumn,aps,showpacs,floatfix]{revtex4}
\usepackage[T1]{fontenc}
\usepackage[latin1]{inputenc}
\usepackage{graphics}

\makeatletter

\newcommand{\LyX}{L\kern-.1667em\lower.25em\hbox{Y}\kern-.125emX\spacefactor1000}

\makeatother

\begin{document}
\title{
Solving the riddle of the bright mismatches:\\
hybridization in oligonucleotide arrays 
}
\author{ Felix Naef and Marcelo O. Magnasco }
\affiliation{Rockefeller University, 1230 York Avenue, New York, U.S.A. }

\begin{abstract}
HDONA 
technology is predicated on two ideas. First, the differential between  high-affinity 
(perfect match, PM) and lower-affinity (mismatch, MM) probes is used to 
minimize cross-hybridization \cite{r2,r3}. Second, several short probes along the transcript 
are combined, introducing redundancy. Both ideas have shown problems in 
practice: MMs are often brighter than PMs, and it is hard to combine the pairs 
because their brightness often spans decades \cite{r4,r5,r6}. Previous analysis suggested these 
problems were sequence-related; publication of the probe sequences has permitted 
us an in-depth study of this issue. Our results suggest that fluorescently labeling 
the nucleotides interferes with mRNA binding, causing a catch-22 since, to be 
detected, the target mRNA must both glow and stick to its probe: without labels it 
cannot be seen even if bound, while with too many it won't bind. We show that this 
conflict causes much of the complexity of HDONA raw data, suggesting that an 
accurate physical understanding of hybridization by incorporating sequence 
information is necessary to perfect microarray analysis.
\end{abstract}
\maketitle

There are two widespread technologies in use today for performing large-scale 
mRNA hybridization experiments: spotted arrays and high-density oligonucleotide
arrays (HDONAs, a.k.a. GeneChip®) \cite{r1}. Such experiments have become 
popular for assessing global changes in gene expression patterns; they 
may be used, in a first instance, as screens to identify genes with interesting behaviour 
on an individual basis; but they also hold the promise to unravel some aspects of the 
tangled web of transcriptional controls \cite{r7,r8}. Hybridization array signal is intrinsically 
"dirty", resulting from compromise to trade quality for quantity, and analysis algorithms 
therefore need to achieve high levels of noise rejection against the real-world noise 
observed in the experiments. There is thus a clear need for the early stage algorithms 
that translate the patterns of light and dark recorded by a laser beam into numbers 
estimating mRNA concentrations to perform optimally. Any inaccuracies introduced at 
that level, i.e. loss of signal or false positive assignments cannot be recovered 
thereafter. In the case of spotted arrays, it seems there is little to do beyond better image 
analysis; HDONAs however have typically between 20 and 40 probes per transcript, 
and a function converting those 20-40 numbers into one number has to be supplied \cite{r4,r9}. 
As we show below, this task is not trivial, owing to the complex nature of mRNA 
hybridization and fluorescence detection in this system. HDONA probes are 25-base 
oligonucleotides grown photolithographically onto a glass surface; about a million 
different such probes can be synthesized on one chip at current densities. Because 25-
mers can exhibit considerable cross-hybridization to a complex background, the system 
was built on two layers. A "differential signal" approach performs a first level of 
rejection of spurious signal, by computing the difference between the brightness of a 
PM probe complimentary to a 25-mer in the target RNA, and a MM probe in which the 
middle nucleotide has been changed to its complement. From the thermodynamics of 
DNA-RNA hybrids in solution \cite{r10} it was expected that the PM probe should have a 
higher affinity for the specific target than the MM probe, while cross-hybridization 
should be roughly equal for both. Second, redundancy was introduced by using several 
probe pairs corresponding to distinct 25-mers along the length of the transcript (see 
Figure 1). 

But these ideas do not translate that easily from hybridization in solution to 
HDONAs. An issue long noticed was the large number of probe pairs for which the 
single mismatch brightness was higher than the perfect match up to a third of all probe 
pairs in some chip models 6. This was easy to notice since early versions of the default 
analysis software would not take this matter into account, and therefore some gene 
concentrations were reported as negative. Why this would happen has been the cause of 
much speculation. A two-dimensional plot of PMs vs. their MMs shows that their joint 
probability distribution appears to have two branches, and it was suggested that 
sequence specific effects are playing a crucial role 6. But in the absence of sequence 
information for the probe pairs, this couldn't be verified. However, Affymetrix has 
recently released the necessary data for addressing the problem explicitly. 

\begin{figure}[htb]
\vspace{0.3cm}
{\centering \resizebox*{0.99\columnwidth}{!}{\includegraphics{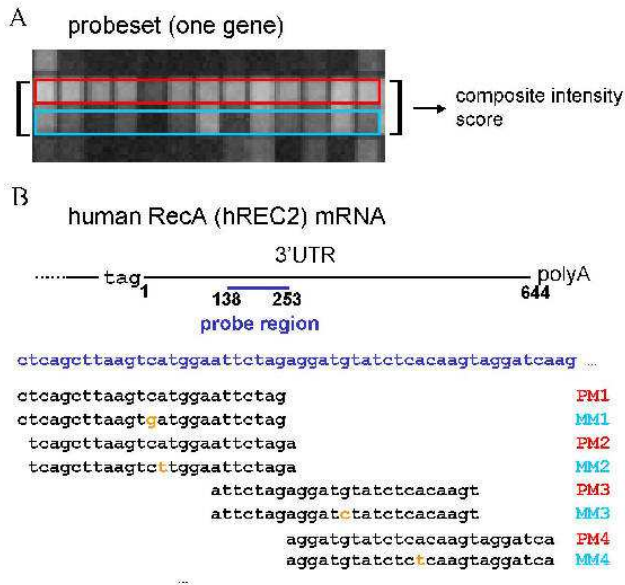}} \par}
\vspace{0.3cm}
\caption{
Probeset design. A: the raw scan of a typical probeset, with the PM 
(respectively MM) on the top (bottom) row. The large variability in probe 
brightness is clearly visible. B: Arrangement of probe sequences along the 
target transcript for the human recA gene in the HG-U95A array; both probing 
the 3'UTR region and the overlap between probes is usual. 
}\end{figure}

\begin{figure}[htb]
\vspace{0.3cm}
{\centering \resizebox*{0.99\columnwidth}{!}{\includegraphics{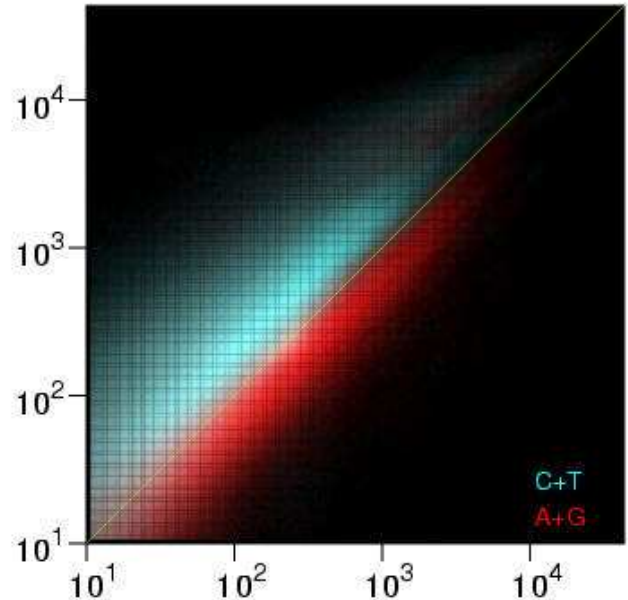}} \par}
\vspace{0.3cm}
\caption{PM vs. MM histogram from 86 human HG-U95A arrays. The joint 
probability distribution for PM and MM shows strong sequence specificity. In this 
diagram, all (PM,MM) pairs in a dataset were used to construct a two-
dimensional histogram---it contains too many points for a scattergram. Pairs 
whose PM middle letter is a pyrimidine (C or T) are shown in cyan, and purines 
(A or G) in red. 33\% of all probe pairs are below the PM=MM diagonal; 95\% of 
these have a purine as their middle letter. 
}\end{figure}

\begin{figure}[htb]
\vspace{0.3cm}
{\centering \resizebox*{0.80\columnwidth}{!}{\includegraphics{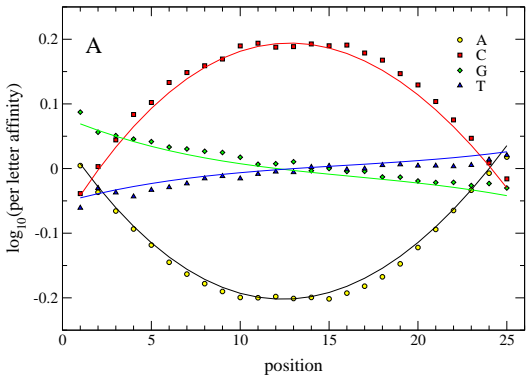}} \par}
\vspace{0.3cm}
\caption{
Sequence specificity of brightness in the PM probes. PM probes from 
the same data as in Figure 2 were fit for as follows: the logarithms of the 
brightnessese divided by a surrogate of concentration (median of all PM's in a 
probeset) were fit (multiple linear regression) to the probe sequence 
composition. At the coarsest level, we fit the data to 100 (4*25) binary variables 
describing the presence or absence of an A, C, G or T at each of the 25 
positions. The resulting site-specific affinities are shown as symbols; position 1 
corresponds to the first base on the glass side. The smoothness of the curves 
permit polynomial fits with much fewer parameters. The solid lines show results 
where the position dependence is modelled as cubic polynomials: we used 13 
(4 parameters * 3 independent letters + offset) variables to fit 17 million data 
points (r2=0.44, F=1071045, p<10-16). The vertical scale is the expected log10 
affinity due to a single letter---thus changing an A for a C at the middle site 
causes the probe to be brighter, on average, by 100.4~250
accumulation of these large sequence affinities results in the exponentially 
broad distribution of measured brightnesses. Notice also the prominent edge 
effects, presumably due to breathing of the duplex. The asymmetry indicates 
effects due both to attachment to the glass and fabrication-specific effects. 
}\end{figure}

\begin{figure}[htb]
\vspace{0.3cm}
{\centering \resizebox*{0.80\columnwidth}{!}{\includegraphics{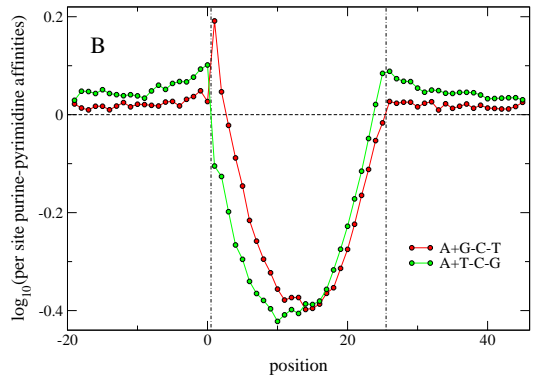}} \par}
\vspace{0.3cm}
\caption{Reduction in brightness due to labeled U and C's. Here fits have been 
extended to also include sequence information from 20 flanking bases on each 
end of the probe. The asymmetry of (A, T) and (G, C) affinities in Figure 3 can 
be explained because only A-U and G-C bonds carry labels (purines U and C 
on the mRNA are labeled). Notice the nearly equal magnitudes of the reduction 
in both type of bonds, additionally, one can observe the change in sign at the 
boundaries of the probes, reflecting the fact that carrying labels outside the 
probe region tends to contribute positively to the brightness, while carrying 
labels inside the probe region is unfavourable because labels interfere with 
binding.
}\end{figure}

We show in Figure 2 joint probability distributions of PMs and MMs, obtained by 
taking every probe pair in a large set of experiments, and binning them to obtain two-
dimensional histograms. We did this twice, computing two separate probability 
distributions which we then superimposed: in red, the distribution for all probe pairs 
whose 13th letter is a purine, and in cyan those whose 13th letter is a pyrimidine. The 
plot clearly shows two very distinct branches in two colors, which correspond to the 
basic physical distinction between the shapes of the bases: purines are large, double 
ringed nucleotides while pyrimidines are smaller single ringed ones. This underscores 
that by replacing the middle letter of the PM to its complementary base, the situation on 
the MM probe is that the middle letter always faces itself, leading to two quite distinct 
outcomes according to the size of the nucleotide. If the letter is a purine, there is no 
room within an undistorted backbone for two large bases, so this mismatch distorts the 
geometry of the double helix, incurring a large steric and stacking cost. But if the letter 
is a pyrimidine, there is room to spare, and the bases just dangle. The only energy lost is 
that of the hydrogen bonds. 

So the existence of two branches agrees with basic hybridization physics, but it 
still does not explain why the MMs are actually brighter than the PMs in many 
sequences with a purine middle letter. To understand this we will perform a finer level 
of analysis, concentrating momentarily only on the PM sequences. It has been pointed 
out that the PMs within a probeset are very broadly distributed, typically spanning two 
decades or more. We can try to observe whether this breadth is similarly sequence-
dependent, by fitting the brightness B of PM probes (divided by the estimated RNA 
concentration [RNA]) against their own sequence composition:
$$ \log\left(B / [RNA] \right) = \sum_{sp} L_{sp} A_{sp} $$
where $s$ is the letter index (ACGT) and $p$ the position (1-25) on the 25-mer; $L$ is a 
Boolean variable equal to 1 if the symbol $p$ equals $s$, and thus $A$ is a per-site affinity. 
More accurate models would include stacking energies by looking at consecutive letters 
(bonds); while this contribution is important for hybridization experiments in solution 
\cite{r11,r12}, we found that it does not improve the fit substantially. On the other hand, we were 
surprised to discover that the key improvement comes from introducing position 
dependent affinities, as opposed to affinities that would depend only on the total number 
of occurrences of each letter. The fitted per-site affinities are shown in Figure 3, note 
the strength of letter specific contributions: changing an A to a C in the middle of the 
sequence would change the brightness of the probe by 250
on mouse, drosophila, and yeast arrays lead to virtually identical affinity curves as those 
shown in Figure 3. Besides providing insight into physical aspects of hybridization, the 
fitted affinities bear an important practical value as they permit to effectively reduce the 
breadth of the probeset brightnesses, therefore improving the signal-to-noise ratio of 
probeset averages (used for instance in absolute concentration estimates). In numbers, 
the variance in 96
by the fit is subtracted, and the reduction is larger than a factor of 2 for 65
probesets. 
An interesting aspect of the above fits is the asymmetry of A vs. T (and G vs. C) 
affinities, which is shown more clearly in Figure 4. 

The obvious culprits for this effect 
are the labels, namely, the standard protocol recommended by Affymetrix entails 
labeling the cRNA with biotinilated nucleotides more specifically, U and C, the 
pyrimidines. This suggests a rather simple explanation, namely, that the biotinilated 
bases somehow impede the binding; the effect diminishing to zero toward the probe 
edges, where the double strand breathes enough to be able to accommodate the linkers, 
and being maximal near the center, where the largest disruption would be effected. This 
would cause a catch-22 in terms of obtaining the maximal fluorescence: if a sequence 
has too few bases that can be labeled, it will not shine, even if it binds strongly, while if 
it has too many labels it will not shine because it does not bind. But this catch-22 has a 
curious loophole: the optimal region to have the fluorophores should then be outside the 
25-mer: since the cRNA being hybridized is usually longer. Figure 4 confirms this: 
when including the contribution to brightness from sequence composition outside the 
25-mer we find the pyrimidine contribution to be strictly positive.

Interference with binding by the biotinilated bases also solves the MM>PM riddle. 
As we saw before, a purine in the middle of the PM implies a gap between the two 
nucleotides on the MM probe; thus one could conjecture that this gap permits the linker 
between nucleotide and biotin not to interfere with the binding. This conjecture is 
quantitatively compatible with the data: according to Figure 4, the energetic penalty for 
a pyrimidine in the middle of the sequence is 0.2 in log10 units (about 0.5 kBT), which is 
comparable to (and bigger than) the excess brightness of the MMs in the purine (red) 
lobe of Figure 2. Indeed, the median excess brightness of the MM for the red probes is 
0.1 in log10 units. In other words, when considering the effective contribution of a 
middle bond to brightness, a G-C* bond on the PM probe is dimmer than a C-C* bond 
on the MM, which in turn is dimmer than a C-G bond on the PM. Here * denotes a 
labeled nucleotide on the mRNA strand.

A microarray experiment carried out for a biological study provides nonetheless a 
quarter of a million measurements in hybridization physics. This information may be 
used to probe and understand the physics of the device, and indeed if an accurate 
enough picture emerges, it shall lead to substantial improvements in data quality. We 
have shown how the basic physics of the detection process in HDONAs percolates into 
the statistics, resulting in statistical anomalies affecting the data thereafter, and which 
need to be taken into account in order to optimize the experiments. Microarrays are one 
out of many high-throughput techniques being developed and brought to bear in 
important problems in Biology today. While it is usually emphasized that they pose 
similar analytical challenges in terms of pattern discovery, mining and visualization, our 
work exemplifies that in order to reach a level where analysis can be abstracted to such 
heights, one should be positive to understand in some detail the physics of the 
instrument and how it affects the data. 
We thank Herman Wijnen, Edward Yang, Nila Patil, Coleen Hacker and Adam Claridge-Chang for 
helpful discussions. Current address (MM): The Abdus Salam International Centre for Theoretical 
Physics, Strada Costiera 11, Trieste I-34100.


\begin{thebibliography}{}
\bibitem{r1} D. J. Lockhart and E. A. Winzeler, Nature 405, 827 (2000).
\bibitem{r2} M. Chee, R. Yang, E. Hubbell, et al., Science 274, 610 (1996).
\bibitem{r3} R. J. Lipshutz, S. P. Fodor, T. R. Gingeras, et al., Nat Genet 21, 20 (1999).
\bibitem{r4} C. Li and W. H. Wong, Proc Natl Acad Sci U S A 98, 31 (2001).
\bibitem{r5} E. Chudin, R. Walker, A. Kosaka, et al., Genome Biol 3, RESEARCH0005. (2002).
\bibitem{r6} F. Naef, D. A. Lim, N. Patil, et al., Phys Rev E 040902 (2002).
\bibitem{r7} D. K. Gifford, Science 293, 2049 (2001).
\bibitem{r8} N. Banerjee and M. Q. Zhang, Curr Opin Microbiol 5, 313 (2002).
\bibitem{r9} F. Naef, C. R. Hacker, N. Patil, et al., Genome Biol 3, RESEARCH0018. (2002).
\bibitem{r10} N. Sugimoto, S. Nakano, M. Katoh, et al., Biochemistry 34, 11211 (1995).
\bibitem{r11} G. Vesnaver and K. J. Breslauer, Proc Natl Acad Sci U S A 88, 3569 (1991).
\bibitem{r12} N. L. Goddard, G. Bonnet, O. Krichevsky, et al., Phys Rev Lett 85, 2400 (2000).
\end{thebibliography}
\end{document}